\documentstyle[12pt,epsf]{article}
\newlength{\dinwidth}
\newlength{\dinmargin}
\begin{document}

\font\blackboard=msbm10 at 12pt
%%%%%%%%%%%%%%%%%%%%%%%%%%%%%%%
 \font\blackboards=msbm7
 \font\blackboardss=msbm5
 \newfam\black
 \textfont\black=\blackboard
 \scriptfont\black=\blackboards
 \scriptscriptfont\black=\blackboardss
 \def\bb#1{{\fam\black\relax#1}}
\newcommand{\z}{{\bb Z}}
\newcommand{\r}{{\bb R}}
\newcommand{\bc}{{\bb C}}
\newcommand{\be}{\begin{equation}}
\newcommand{\ee}{\end{equation}}
\newcommand{\ber}{\begin{eqnarray}}
\newcommand{\eer}{\end{eqnarray}}
\newcommand{\lp}{\left(}
\newcommand{\rp}{\right)}
\newcommand{\lk}{\left\{}
\newcommand{\rk}{\right\}}
\newcommand{\lc}{\left[}
\newcommand{\rc}{\right]}
\newcommand{\ls}{\alpha'}
\newcommand{\cm}{\hspace{1cm}}

\baselineskip18pt

\thispagestyle{empty}

\begin{flushright}
\begin{tabular}{l}
FFUOV-00/02\\
{\tt hep-th/0002052}\\
%\today
\end{tabular}
\end{flushright}

\vspace*{2cm}
{\vbox{\centerline{{\Large{\bf HOLOGRAPHY AND T-DUALITY}
}}}}

\vskip30pt

\centerline{Marco Laucelli Meana and Jes\'{u}s Puente
Pe\~{n}alba 
\footnote{E-mail address:
     laucelli, jesus@string1.ciencias.uniovi.es}}

\vskip6pt
\centerline{{\it Dpto. de F\'{\i}sica, Universidad de Oviedo}}
\centerline{{\it Avda. Calvo Sotelo 18}}
\centerline{{\it E-33007 Oviedo, Asturias, Spain}}

\vskip .5in

\begin{center}
{\bf Abstract}
\end{center}
We use the AdS/SYM correspondence to study the relevant effects of  compactified
dimensions on the D-brane dynamics. We present a detailed picture of the T-duality
transition between branes in type IIA and type IIB supergravity. An analysis 
of the renormalization scheme coming from the expectation values of background
fields and the role of Wilson lines in it is given. We finally explore finite 
size effects and T-duality maps on the description of Wilson loops by supergravity.

%\vskip 18 pt

\newpage

\section{Introduction}

The celebrated AdS/CFT conjecture \cite{malda,poly, witten1} opened, 
two years ago, an important path to study strongly coupled gauge theories
from an unexpected supergravity framework. With this new technology, the string theorists 
have been able to describe a variety of properties of gauge theories unreachable
from the field theory point of view (see \cite{rev}). All this work was supported by the 
conformal  invariance of the ${\cal N}=4$ SYM on the D$3$-brane and represented 
in supergravity  by the conformal nature of the $AdS_5\times S^5$ of the 
corresponding background. This symmetry is a strong constraint on the systems that 
has allowed a good description of it.

The question we will approach in this paper is in the direction of the extension of
the Maldacena's conjecture to non-conformal branes. Some work on the subject has been
done \cite{son}. In particular we will focus on the compatibility of a string symmetry, 
T-duality, with the holographic conjecture. There have been some approaches to the problem 
\cite{silva,korea} and it seems that, at least, part of the phase space of branes on 
tori has been constructed \cite{martinec,barbon}. 

Our first interest in this paper is the clarification of such results. We will argue that 
T-duality does not affect the holographic description of D-brane dynamics, and we could 
map from one brane system to the other, any physical property of the system. We will
specially focus on the renormalization of classical parameters, and discuss how it must
be understood.

We will explore the corrections of the dynamics due to the presence of compact dimensions. 
This analysis will allow us to present of a detailed description of different parts of
the phase space. In particular we describe an intermediate phase between two T-dual systems
that is exclusively produced by finite size effects.

Finally we will explore the dynamics of Wilson loops in this kind of backgrounds. The 
analysis of this system will be done with the same philosophy of the previous sections.
We will study two different quark-antiquark configurations and see how T-duality and 
finite size effects change their properties.

\section{The background solution.}

  We are interested in obtaining the complete background solution for
wrapped D-branes and their T-duals, that is, unwrapped D-branes moving in a
transverse space that is compactified in at least one direction.

  We shall compute one (the latter) and deduce the other with the help of
the Buscher rules for T-duality\cite{tdual}. For definiteness, let us concentrate in the
case of D2-branes with a transverse space like $\r^6\times S^1$. It is easier
if we adopt the multi-centered image, which consists in considering the
circle as the whole $\r$ space with an identification $x \sim x+2 \pi R$.
This means that the compactified metric is equivalent to the one induced by
an infinite number of parallel and equally-spaced D2-branes. Of course, the
solutions cannot be linearly added because Supergravity is not at all a
linear theory. However, happily enough, we can write everything in terms of
a function, usually called harmonic function, that does behave linearly \cite{horo}. All
physical fields, in particular, the metric and the dilaton, are non-linear
functionals of it.

   The general $p$-brane solution is 

\be
ds^2=f^{-1/2}(-dt^2+dx_i^2)+f^{1/2} d y_j^2
\ee

with

\be
e^\phi=g_s f^\frac{3-p}{4}
\ee
where $i$ is the index of the directions parallel to the brane, $f$ is the
harmonic function that depends on the transverse coordinates and on the
radius of the compact direction. This function $f$ obeys a Laplace equation
whose solution, for a group of $N$ superposed D2-branes is\cite{horo}
\be
f(r)=1+\frac{g_s N l_s^5 d_2}{r^5}
\ee
where $d_2=\frac{(2\pi)^5\Gamma(7/2)}{10 \pi^{7/2}}$. The multi-centered
solution is 
\be
f(r)=1+\sum_{n=-\infty}^\infty 
\frac{g_s N l_s^5 d_2}{\lc r_\bot^2+ \lp r_\|+2 \pi n R \rp^2 \rc^{5/2}}
\ee

   The subscripts separate the coordinate that parameterizes the circle from
the other transverse ones. Now we take the decoupling limit \cite{malda}to see if 
we get
something similar to an AdS space. We must define $u_\bot:=r_\bot l_s^{-2}$,
$u_\|:= r_\| l_s^{-2}$, and also ${\cal R}:=R l_s^{-2}$. We take $l_s$ and
all other lengths to zero but keeping the $u$ and ${\cal R}$ variables
finite. The first two follow the usual infrared limit that is necessary to
decouple the open strings attached to the branes from the closed strings. In
the supergravity, this separates the fields moving in the Minkowskian
geometry very far from the horizon from the fields very near it. The meaning
of the definition of ${\cal R}$ is not the same; $r$ and $u$ are variables
and the limit is a restriction of the values, on the other hand, the radius
is a constant and what we are doing is to choose the value of $R$ in such a
way that ${\cal R}$ is finite (and much smaller than $l_s^{-1}$). We will
later explain the reason for this choice.

   The only relevant magnitudes in the new geometry will be the ones
defined above. We obtain

\be 
f(u_\bot, u_\|,{\cal R})=\frac{g_s N d_2}{l_s^5} \sum_{n=-\infty}^\infty 
\lc u_\bot^2+ \lp u_\|+2 \pi n {\cal R} \rp^2 \rc^{-5/2}        
\ee

  It is useful to use spherical coordinates in the six-dimensional transverse
space and angular ones for the compact circle. This way ${\cal R}$ can
explicitly appear in the metric. That way the solution becomes

\be
ds^2=f^{-1/2}(-dt^2+dx_i^2)+f^{1/2}{\cal R}^2 d\theta^2+ f^{1/2} du_\bot^2+
u_\bot^2 f^{1/2} d\Omega_5^2
\label{d2-metric}
\ee

  $\theta=u_\|/{\cal R}$ is dimensionless and has periodicity $2\pi$. The
three longitudinal coordinates and $u_\bot$ form a kind of AdS space over
which a sphere and a circle, both with variable radii, are fibered. The
T-dual metric is obtained with the substitution
\be
f^{1/2}{\cal R}^2 d\theta^2 \longrightarrow f^{-1/2}{\cal R}^{-2}\alpha'^{-2} 
d\psi^2
\ee

   The way we have dealt with the solution does not seem to be symmetric
with respect to the duality. We have constructed the solution deforming the
uncompactified solution of the D2-branes, but in the limit ${\cal R}
\rightarrow 0$ we should recover the D3-brane case. There should, then,
exist an expression that deforms the D3-brane solution adding the
contribution of the closed strings wound around the compact Dirichlet
direction, instead of the contribution of the mirror images, as we have
done. It can be obtained by a Poisson resummation of the function $f$.

\ber
f(u_\bot, \theta,{\cal R})=\frac{g_s N d_2}{l_s^5} 
\sum_{n=-\infty}^\infty \lc u_\bot^2+ {\cal R}^2\lp \theta
+2 \pi n \rp^2  \rc^{-5/2}= \nonumber \\       
=\frac{g_s N d_3}{l_s^5 {\cal R}} u_\bot^{-4}+ 
\sum_{n=1}^\infty \frac{g_s N d_3}{l_s^5 {\cal R}^3} u_\bot^{-2}
\cos\lp n \theta \rp K_2\lp\left|\frac{n u_\bot}{{\cal
R}}\right|\rp
\label{series}
\eer

   where $d_3=4\pi$. The first term of the sum is the solution for the
D3-brane with coupling $g_s'=g_s/(l_s {\cal R})$. Notice that the natural
transition point is $u_\bot \sim {\cal R}$. By `transition point' we mean
the value of the variables where the zero mode (first term) of one of the
series begins to be a bad approximation as more and more terms need to be
added to get a certain accuracy, and the zero mode of the other series
begins to be a good approximation.

   The physical meaning of this resummation is something similar to the
Gregory-Laflamme \cite{martinec,gl} localization transition of black holes. Here we 
do not have
a horizon, and therefore, neither have we a definite size. However, at a
certain transverse distance (in $u_\bot$) of the center, the D2-brane
solution has a width in the longitudinal direction ($u_\|$) that can define
a typical size, precisely of order $\cal R$. When the distance to the center
is larger than that, the metric is basically constant with respect to
$u_\|$. In the picture where the circle is emulated by a one-dimensional
lattice, the observer beyond the point $u_\bot=\cal R$ sees a nearly
continuous distribution of branes in the compact direction. Any computation
obtained as a sum of contributions of each individual D2-brane has to be
resummed.

   We expect to relate this soft transition to T-duality so it is somewhat
surprising that $\alpha'$ does not appear in the condition. The explanation
is as follows. There are two different reasons to prefer a theory instead of
its T-dual. In the case of the closed strings, there are two series of
energy levels in the spectrum, windings and momenta. If the winding modes
are much lighter than the momenta, it is necessary to resum the perturbative
series and therefore, it is more useful to use the T-dual theory. When
$R=\sqrt{\alpha'}$, both modes have the same energy and that is why that is
the limit of usefulness of each theory. On the other
hand, when only open strings are involved, windings and momentum cannot
coexist. Only when there is another length (or energy) scale like the
separation of branes is there a reason to prefer one of the two theories. If
the gaps in the discrete spectrum are much smaller than the separation of
the branes many of them will contribute to a typical process and it is
better to use the theory where they are momenta and can be easily
integrated. Otherwise, one should use the other one. The range where the
transition happens is around $R=Y$, if $Y$ is the separation between the
branes.

   This second possibility is the one we have found in this section.

\section{The gauge theories.}

   Let us now see which is the field theory that is described by this
geometry. It is the one that appears when we take the same limit in the open
string model of the same group of D-branes. The modes of the string that do
not decouple are the massless ones, whose dynamics is governed by the
SYM$_{2+1}$; and the strings that wind a finite number of times around the
compact direction before both ends attach on the D2-brane. Again, we should
use the multi-centered image, which is the clearest. The field theory is a
$U(N\times \infty)$ SYM in $2+1$ dimensions. $N$ is the number of parallel and
superposed D2-branes, it must be large in order for the conjecture to work
well. The infinity that multiplies that $N$ can be substituted by any number
large enough to describe all the windings of the system. The scalar with
index in the compact direction takes the following expectation value
\be
X^i=blockdiag\{...,-Id_{N\times N},0_{N\times N},Id_{N\times N},... \} 
\ee
that breaks the gauge symmetry to the subgroup $U(N)^\infty$.

   There is a T-dual description of this system that interchanges windings
and momenta, which are, in this case Higgs' masses and Kaluza-Klein momenta.
It is the SYM$_{3+1}$ compactified in a circle of radius $\tilde{R}$ (that
equals $\alpha'/R={\cal R}^{-1}$). The isomorphism identifies the fields of
the $N\times N$ boxes which are, say, $M$ places away from the diagonal with
the $N\times N$ fields that complete the adjoint representation of the
SYM$_{3+1}$ with KK momentum $M/\tilde{R}$ (=$M {\cal R}$). There are two
ways to write the action of this theory: as a three-dimensional theory with
a compact direction or as a two-dimensional theory with an infinite number
of fields of relatively integer masses. When the radius $\tilde{R}$ is
small, or better, when the typical energy of the experiment is smaller than
the first KK level, the two-dimensional representation is more suitable
while in the opposite case, one should use the three-dimensional action.

   Both actions reproduce the same physics and, therefore, give the same
results. Renormalized parameters are functions of $f$, for example
\be
g_{YM}=e^{\phi/2}=g_s^{1/2}f^{1/8}
\ee
The interpretation 
of the sum we have written in formula (\ref{series}) is that it is the way the
masses originated by the compactification add different corrections to the 
renormalization of the
two uncompactified theories and interpolate between them in a continuous
manner. There is, however, one subtlety. The interpolation is soft between
the two behaviours of the function $f$, but in the solution of both the
dilaton (the gauge coupling) and the metric there is a discrete jump: in the 
first case, there is
a different power of $f$ and in the second there is an inversion of one of
the components. The meaning of this is that the two different actions that
describe the field theory do it in terms of two kinds of fields. The two-
and three-dimensional fields have different dimensions and therefore do not
scale equally \footnote{As usual \cite{witten1} computing two-point functions of
reduced fields on the transverse sphere one finds that the weights of
 three-dimensional fields correspond to masses on the $3$-brane background. 
For two-dimensional fields we should reduce fields of type IIA supergravity on 
$S^1\times S^5$, and their masses should account for their weight. In this case 
both the supergravity and gauge theory fields hold in representations of 
$SU(4)\times U(1)$ R-symmetry group.} .In particular, the gauge coupling of the 
three-dimensional
theory is dimensionless and, classically, it is independent of the scale; on
the other hand, in the theory with two dimensions, the coupling is a length
to the $1/2$ power and the dimensionless coupling ($g_{YM} \sqrt{U}$)
depends on the scale. In principle, both descriptions are not only
equivalent, but, in fact, are exactly the same. In both cases loops take the
form of sums in a discrete number which we can call momentum or winding, but
it is the same thing. So the picture is that the effective coupling when,
for example, the energy is very low, is that of the D2-brane theory. If we
are interested in higher energies, we reach a region in phase space where
there is an increasing number of addends in the function $f$ that contribute
significantly, and when they are neatly higher than $\cal R$, the series can
be resummed to obtain a new good zero mode. In terms of fields, this
resummation can be seen as the redefinition of two-dimensional fields in
terms of three-dimensional ones:

\be
A^i(x_3,x_j)=\sum_{n=-\infty}^\infty X_n(x_j) e^{-i n x_3/\tilde{R}}
\ee

  This change of variables is not trivial at all when one considers the
renormalized theories because $R$, as any other parameter appearing in the
Lagrangian acquires a dependence on the scale $u$, that we should introduce
in the Fourier expansion. This is a consequence of the renormalization of
the masses of the KK modes. The redefinition affects the coupling constant
as it does in any dimensional reduction:

\be
(g_{YM})^2=\frac{\lp g_{YM}' \rp^2}{R_{\mbox{ren}}(u)}
\ee 

  This means that the behaviour under renormalization of both theories can
be very different. In our case, using the correspondence, one finds

\be 
u_\bot^{-5/8} \stackrel{u \ll {\cal R}}{\longleftarrow} g_{YM}
\stackrel{u\gg {\cal R}}{\longrightarrow} u_\bot^{-1/2} 
\ee 

while $g_{YM}'$ is independent of the scale. This is due to the fact that the
dependence on the scale of $g_{YM}$ and that of $R_{\mbox{ren}}(u)$
exactly cancel. In fact, the behaviour $g_{YM}\propto u_\bot^{-1/2}$ is
conformal because the dimensionless coupling in two dimensions $g_{YM}
\sqrt{u_\bot}$ is, indeed, independent of the scale.

  It is interesting to notice that in this case the map that relates both
sides of the correspondence is a little more complex. Indeed, the
renormalized magnitudes depend on three parameters ($u_\bot$, $\theta$ and
${\cal R}$) instead of one. From the original correspondence for the
D2-brane case we know that $u_\bot$ is the $\mu$ parameter, the typical
energy of the experiment which we are testing the system with. The reason
why ${\cal R}$ appears in the D2-brane side is that the renormalization that
the supergravity gives uses a mass-dependent scheme. The massive fields give
contributions to the renormalized quantities in such a way that when
energies ($\mu$) are smaller than their mass, they naturally decouple due to
the fact that the terms of the series related to them tend to zero in that
limit. In our case, those terms are Lorentzian-like. In a mass-independent
scheme, the terms are added by hand when necessary. The translation would be

\be
\lc u_\bot^2+ {\cal R}^2\lp \theta+2 \pi n \rp^2 \rc^{-5/2} \longrightarrow 
u_\bot^{-5} {\cal H} \lp u_\bot-{\cal R}\left| \theta+2 \pi n \right| \rp 
\ee
where ${\cal H}$ is a Heaviside step function. Although the function tends
to a constant, not to zero, when the variable is small, it is possible to
neglect it because it is much smaller than the divergent zero mode term. The
variable $\theta$ gives a contribution to the mass because from the point of
view of the field theory, it is the expectation value of the scalar field
that is used as a test; this breaks the gauge symmetry and changes the mass
of some fields.

  In the picture with the D3-branes, there are not any masses but discrete
momenta. We know that in mass-independent schemes, compactification does not
affect the renormalization and no dependence on the radius appears. The
reason is that mass-independent renormalization adds the minimum
counterterms to the Feynman diagrams to cancel the ultraviolet divergences.
Obviously, compactifying does not affect at all the ultraviolet behaviour
because it is a purely infrared phenomenon. However, if one wants to observe
how the infrared effects act over the effective coupling, one should use a
radius-dependent scheme and that is exactly what the supergravity result is.
Indeed, the first term is the one given by the minimum counterterms and the
rest are exponentially suppressed when the energy is larger than the inverse
radius. In this case $\theta$ is related to the Wilson line. The precise
identification is 
\be
A^i=\frac{{\cal R}}{2\pi}\theta
\ee

The reason why it appears as a renormalization parameter is that the
interaction, and therefore the coupling constant is affected by its
presence. Its effects are periodic, that is why they appear as a Fourier
expansion, and depend on the proportion between $u_\bot$ and ${\cal R}$. If
the radius is very small, then the exponential asymptotic behaviour of the
Bessel functions makes the contribution of the Wilson line negligible, but
if the radius is large, then the function $f$ develops a clear maximum in
$\theta=0$ and also does the coupling. This is due to the fact that the Wilson
line breaks the gauge symmetry and gives masses to the intermediate
(off-diagonal) bosons. In the infrared, when $u_\bot$ is much smaller than
${\cal R}$, these masses can forbid any interaction and that is why the
effective coupling can tend to zero and is maximum when the Wilson line is
not present at all ($\theta=0$).

\section{Effects of renormalization on the phase space.}

   As we have written just above, not only the coupling changes with
the `distance' $u$, also the radii of both the five-sphere and the circle
change and it is interesting to analyze its meaning. Results related to this
section are in \cite{son,martinec}.

   Let us start with the circle \footnote{We are very grateful to Mar\'{\i}a Su\'arez
for her help on the topics covered in this section.}. In the supergravity, 
${\cal R}$ is a
parameter related to the value of the radius measured by a faraway
observer situated where the metric is Minkowskian. The limit we have taken
has completely disconnected that observer from the system we want to study
and the physical radius is a function given by
\be
R_{\mbox{phys}}(u_\bot,\theta,{\cal R})=\alpha' {\cal R} 
f^{1/4}(u_\bot,\theta,{\cal R})
\ee
for the D2-brane case and the inverse for the D3-brane one.

   The interpretation of this is direct. In any field theory all parameters,
including radii can be renormalized. The expression above tells us how the
physical, renormalized radius depends on the scale. This can also be seen as
the renormalization of the masses of the Kaluza-Klein modes.

   Let us now discuss the low energy spectrum of the two (IIA and IIB)
T-dual string background solutions that we have. We are interested in
particular in the energies of winding modes and discrete momenta.
Classically, if we choose ${\cal R}=R/l_s^2$ to be an order one magnitude,
then the energy of the IIA windings is also order one ($2\pi m{\cal R}$),
while that of the momenta is large (${\cal R}^{-1} l_s^{-2}$) and decouples.
If we now look at the physical radius, because of the fact that the function
$f$ runs from $0$ to infinity, this is not true in all points of space. In
fact, there is a point where
$R_{\mbox{phys}}(u_\bot,\theta,{\cal R})=l_s$ where momenta begin to be
the lightest modes and must be taken into account while windings decouple.

This is the usual phenomenon that sets the limit of utility of two T-dual
closed string theories, as we have written before. In this case, this soft
transition always happens when $u\gg{\cal R}$, that is, in the far
ultraviolet of the Yang-Mills theories. Explicitly
\be
R_{\mbox{phys}}(u,\theta,{\cal R})=l_s \hspace{1cm} \mbox{if}
\hspace{1cm} u_\bot \simeq d_3^{1/4}(g_s N)^{1/4} l_s^{-1/4} {\cal R}^{3/4}
\label{transition}
\ee
That limiting value for $u$ is much larger than $\cal R$ because 
$(g_s N)/l_s$ tends to infinity in order for the conjecture to work. For this
reason, this transition, always occurs when $f$ can be well approximated by
\be
f(u_\bot,\theta,{\cal R})=\frac{d_3 g_s N}{l_s^5 {\cal R}u_\bot^4}
\ee
that is the harmonic function of a D3-brane in an open space. When $u_\bot$
is beyond both transition points, in the ultraviolet, the D3-brane solution
is exact and there is not any finite size effect. When it is smaller ($u \ll
\cal R$), it is the D2-brane solution the one that describes well the
problem. When
\be
{\cal R} < u < (g_s N)^{1/4} l_s^{-1/4} {\cal R}^{3/4}
\ee
there is an intermediate case. One can use the supergravity solution of the
D3-brane, but size of the circle is such that the momenta are light
enough to be considered continuous but windings (which are string
corrections) are even lighter and have to be added. On the other hand, one
could use T-duality and find the solution in terms of D2-branes; then
windings are heavier than momenta, as they should, but still, the harmonic
function must be resummed. In the picture where the circle is seen as
$\r/\z$, the solution is that of a continuous set of D2-branes that fill one
of the perpendicular directions.

   The question now is how can we interpret this transition from the point
of view of the Yang-Mills theories. The guides are group theory and the
identification of $R_{\mbox{phys}}(u)$ as the renormalized value of $\cal
R$. The supergravity fields that have discrete momentum in the IIB theory
(or winding in the IIA one) clearly correspond to the Kaluza-Klein modes of
the SYM$_{3+1}$ (or the massive off-diagonal modes in the $U(N\times\infty)$
SYM$_{2+1}$). The relation is the charge under the $U(1)$ symmetry of the
circle.

   Compactifying a Yang-Mills theory in a circle has several consequences:
the momenta get discrete and the component of the gauge field polarized in
that direction acquires a periodicity $A^i \sim A^i+\frac{1}{R}$, besides
that field can have a vacuum expectation value (Wilson line). The
periodicity affects the momentum of the field (the electric field in that
direction) making it discrete with gap $2\pi R$. The electric field has
energy proportional to $\vec{E}^{\,2}$. The renormalization of $\cal R$
makes this energy spectrum identical to that of the windings around the
compact direction (in the IIB 3-brane supergravity). We can be more precise.
Let us design an experiment with an observer in the IIA supergravity theory
moving in the direction of the circle, that is one D2-brane, separated from
the rest, falling towards the central bundle of $N$ D2-branes. From the
point of view of the $U(N+1)$ SYM$_{2+1}$, the movement is described by a
time-dependent vacuum expectation value of the scalar that breaks the
$U(N+1)$ into $U(N)\times U(1)$. This represents the change of the
coordinate of the D2-brane as it moves. With the usual identification, that
velocity is related in the $U(N+1)$ SYM$_{3+1}$ to the value of a background
electric field. As we have taken our observer to be a D2-brane, its movement
is Galilean and that is why the dispersion relation energy-electric field is
quadratic.

   With this we learn a bit about the meaning of the closed string T-duality
for the correspondence. If in usual string theory it could be seen as a
competence between momenta and windings, here it can also be seen as a
competence between two different spectra. In perturbative gauge theories
there are two kinds of excitations: the perturbative fields and the solitons
that represent changes in the vacuum. In our case the vacuum is
parameterized by the expectation values of the scalars. Usually, unable to
do exact computations, one uses the Born-Oppenheimer approximation and
consider the solitons as `slow modes' and the perturbative fields as `fast
modes'. Their dynamics have so different frequencies or time scales that
completely decouple and one can solve them independently. In our case, the
D-branes are so heavy that their movements and those of the open strings are
independent. In our case we have found that when a Yang-Mills theory (at
least of the kind we are studying) is compactified, depending on the value
of the energy in terms of the radius and the other parameters ($g_s$ and
$N$) both types of modes (polarized in the compact direction) can
interchange the role of being the lightest. When the energy is high, the
slow modes are very heavy while the fast ones are light, but below the
transition point written in (\ref{transition}), the situation is the
opposite.

   To describe the intermediate phase we can either use the $U(N)$
SYM$_{3+1}$ with one compact direction, light KK modes but even lighter
electric fields or (better we would say) the $U(N\times \infty)$ SYM$_{2+1}$
with the gauge symmetry slightly broken by some light off-diagonal modes.

  Once we have dealt with the radius, let us see now what happens to the 
couplings. They are
\be
g_{IIB}(u_\bot,u_\|,{\cal R})=g_s^{IIB}\hspace{1cm}\mbox{and}\hspace{1cm}
g_{IIA}(u_\bot,u_\|,{\cal R})=f^{1/4}g_s^{IIA}=f^{1/4}g_s^{IIB} {\cal R} l_s
\ee

  As the IIB string theory is S-selfdual, we can choose $g_s^{IIB}$ to be
smaller than one. However, we cannot prevent that $g_{IIA}$ become larger
than one in some region in the target space, when
\be
u < \lp g_s^{IIA} \rp^{5/4} N^{1/4} {\cal R}^{-1/4} l_s^{-5/4}.
\ee
There, it is necessary to consider corrections from M theory.

   Another limit is imposed by the variable radius of the sphere. When it is
too small, the curvature can exceed the string scale and then it is also
necessary to add corrections. This is not different from the usual uncompactified cases so
we shall not discuss it more.

\begin{figure}
%
%\vspace*{2 cm}
%
%%Begin InstantTeX Picture
\let\picnaturalsize=N
\def\picsize{4in}
\def\picfilename{phase.ps}
%If you do not have the picture file add:
%\let\nopictures=Y   
%to the beginning of the file.
\ifx\nopictures Y\else{\ifx\epsfloaded Y\else\input epsf \fi
\let\epsfloaded=Y    
\centerline{\ifx\picnaturalsize N\epsfxsize \picsize\fi
\epsfbox{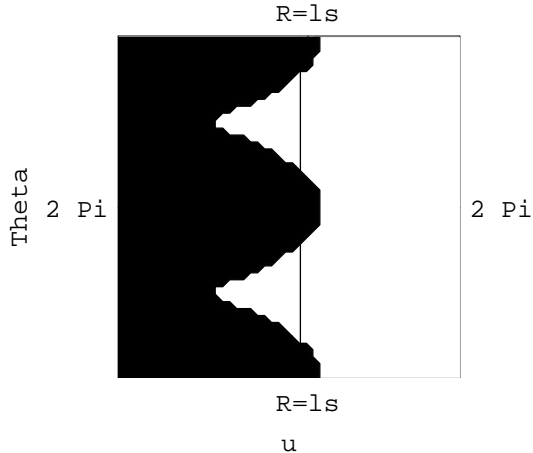}}}\fi
%%End InstantTex Picture   
\caption{{\small T-dual phases in terms of the value of the physical radius. We see how
finite size effect changes the form of the transition line $R_{\mbox{phys}}=l_s$. The
straight line represents the transition without considering finite size effects.}}  
\label{radius}
\end{figure}

   In the picture (\ref{radius}), we show the effect of the Wilson line ($\theta$ parameter)
on the phase space. The line that separates the two phases, the one describable in terms of 
D3-branes and the other in terms of the continuous set of D2-branes, is not a straight line.
The Wilson line extends or reduces each phase depending on its value. More 
to the left, we could draw a second wavy line, parallel to this one, that signalled the 
appearance of an M2-brane phase when the coupling constant is larger than one. The phase 
with D2-branes in an open space ($u_\bot<\cal R$) can exist or not. It does exist if the
M2-brane phase appears precisely when $u_\bot<\cal R$. The line of phases is, therefore: 
wrapped D3-brane, continuous set of D2-branes, D2-brane in open space (not always)
and M2-brane. The straight line drawn in the figure is the result obtained with the D3-brane
neglecting all finite volume effects.

\section{Wilson loops}

In this section we will analyze Wilson loops in theories with a compactified 
dimension. As in the rest of our work we will deal with the D$2$ and D$3$-brane 
systems. We will focus our attention on those features coming from the presence of
a compactified dimension.

Let us remind how we can introduce static quarks in gauge theories living on the 
brane. We begin with $N+1$ branes and we break the $U(N+1)$ gauge theory by taking 
one brane to infinity. The open strings connecting the separated brane to the others 
have infinite mass and transform in the fundamental (antifundamental) of the gauge 
group. These are the external quarks. From the worldvolume point of view these 
sources produce solitonic deformations on the worldvolume because the $N$-branes are 
pulled by the infinitely long open strings attached to the other. 

The shape of this deformation could be represented, on the supergravity side,  by 
the world-sheet of a string ending on the boundary of the background manifold. Using
this prescription we can stablish a concrete correspondence between the 
quark-antiquark potential and the classical action of the string as \cite{maldaloop,rey}
\be
<W(C)>_{Gauge}=A(L)e^{-T E(L)}=<e^{-S}>_{Sugra}
\ee
where $C$ is the Wilson contour and $L$ is the distance from one source to the other.
 The string action is computed as
\be
S=\frac{1}{2\pi \alpha'}\int dx dt\sqrt{det  G_{mn}\partial_{a}X^m\partial_{b}X^n}
\ee
where $G_{mn}$ is the euclidean background metric in the loop directions. 
In order to compute
the classical action we should find the string configuration that minimizes the
world-sheet area. 

We are interested in finite size effects due to the compact dimension. We will 
investigate how to compute the Wilson loop in the D$2$-brane case and 
compare it with the result obtained in the D$3$-brane theory. There are several 
situations to be considered and we will describe the T-duality map for each case. These
 situations are related to different loop geometries.

Suppose  a quark-antiquark pair in 
the D$2$-brane theory separated in a world-volume direction. Remember that in this case
 the boundary theory has a $SO(6)\times U(1)$ R-symmetry that corresponds to rotational
 symmetry  on the supergravity side. This allows us to take quarks on different points
 on the  $S^5\times S^1$ transverse manifold and introduce an angular difference 
between the open strings connecting the separated branes and the $N$ central ones. 

We will begin with the simplest  configuration that corresponds to taking
the angular difference to zero. Using the metrics in (\ref{d2-metric}) we can
see that the classical world-sheet action
 is given by
\be
S=\frac{T}{2 \pi \alpha'}\int dx \sqrt{ G_{tt}G_{xx}+G_{tt}G_{uu} (\partial_x u)^2}
\label{NG}
\ee
where we have integrated the temporal variable assuming a static configuration for the 
Loop.

We can compute the worldsheet area in (\ref{NG}) using a general result as, for example,
in  \cite{sfetsos}. These general  formulas allow us to stablish that if the 
Nambu-Goto action is written as
\be
S=\frac{T}{2 \pi \alpha'}\int dx \sqrt{A(u)(\partial_x u)^2+ \frac{B(u)}{\lambda^4}}
\ee
then the physical magnitudes that we are interested in could be written
\ber
L=2 \lambda^2 \sqrt{B(u_0)}\int_{u_0}^{\infty} du \sqrt{\frac{A(u)}{B(u)(B(u)-B(u_0))}}
\nonumber \\
E=\frac{1}{\alpha'\pi}\int_{u_0}^{\infty} du \left[ \sqrt{\frac{A(u)B(u)}{B(u)-B(u_0)}} 
\right]-
\frac{1}{\alpha' \pi}\int_{u_{min}}^{u_0} du \sqrt{A(u)}
\label{general}
\eer
where $u_0$ is the point given by $u(x=0)=u_0$, and $u_{min}$ is a
geometrical minimal value of $u$ given, for example by the presence of a
singularity on the background manifold.

We can now compute the Wilson loop in the D$2$-brane theory. Using the action in 
(\ref{NG})and expressing it in terms of the functions in (\ref{general}) we can write
\ber
A(u_{\bot})=G_{uu}G_{tt}=\alpha'^2 \nonumber\\
B(u_{\bot},u_{\|},{\cal R})=\lambda^4 G_{xx}G_{tt}= \alpha'^2 
H(u_{\bot},u_{\|},{\cal R})^{-1}
\eer
where $\lambda^4=d_2 g_{YM} N$ and $H(u_{\bot},u_{\|},{\cal R})$ is the sum over 
multicenter solutions in the first line of (\ref{series}). Finally one obtains
\ber
L=2 \lambda^2\int_{u_0}^{\infty} du \frac{H(u,u_{\|},{\cal R})}{\sqrt{H(u_0,u_{\|},
{\cal R})-
H(u,u_{\|},{\cal R})}}
\nonumber \\
E_{qq}=\frac{1}{\pi}\int_{u_0}^{\infty} du \sqrt{\frac{H(u_0,u_{\|},{\cal R})}
{H(u_0,u_{\|},{\cal R})-H(u,u_{\|},{\cal R})}}.
\label{wl2}
\eer

The final scope of our computation should be the resolution of $E$ in terms of $L$. It 
does not seem possible to obtain this result analytically but it is possible to do it 
numerically. We can compute both integrals in terms of the variable $u_0$ and then 
extract a numerical representation of $E=E(L)$. We will recover the analytic result of 
the Wilson loop in the limit  ${\cal R}\rightarrow \infty$, which corresponds to the 
D$2$-brane on a decompactified background.   

The complete T-dual map for this quark-antiquark computation is obtained by computing 
the Wilson Loop from the D$3$-brane point of view. The method is exactly the same. 
In this case we will use the Poisson resummed expression for the harmonic function
which is given in the second line of (\ref{series}). We can write it as
\be
B(u_{\bot})=\lambda^4 f^{-1}=\alpha'^2 u^4_{\bot} g(\tilde{R},u_{\|},u_{\bot})^{-1}
\ee
now $\lambda^4=d_3 N g'_s$. Using this parameterization of the metric and the usual 
procedure we can express, as we did for the D$2$-brane, the energy of the Wilson loop 
and  the separation between quarks in  terms of $u_0=u_{\bot}(x=0)$ 
\ber
L(u_0,{\cal R},u_\|)=2 \lambda^2 u_0^2 \int_{u_0}^{\infty} \frac{du}{u^4}
\frac{g(u)}{\sqrt{g(u_0)-g(u) \lp \frac{u_0}{u}\rp^4}} \nonumber \\
E_{qq}(u_0,{\cal R},u_\|)=\frac{1}{\pi}\int_{u_0}^{\infty} 
\frac{du}{\sqrt{g(u_0)-g(u)
\lp \frac{u_0}{u}\rp^4}}-\frac{1}{\pi}\int_{u_0}^{\infty} du
\label{WL3}
\eer
where the last integral eliminates the usual divergence coming from the quark masses. 
We can recover the standard D$3$-brane result simply taking the radius parameter 
$\tilde{R}$ to infinity then reducing the complete series in $ g(\tilde{R},u_{\|},
u_{\bot})$ to its zero mode. 

The first evidence that  comes out directly from the results in (\ref{wl2}) and 
(\ref{WL3}) is that both solutions are exactly the same, but expressed in T-dual 
variables. This fact simply  reflects that, for this simple configuration of the Wilson 
loop, T-duality does not affect any parameter defining the system. More concretely, 
what we have seen is that the effective metric, used to compute the string worldsheet 
area, remains unchanged under Busher's transformation rules. Consequently, they give 
the same result for the quark-antiquark potentials in D$2$ and  D$3$ brane theories. 
\begin{figure}
\vspace*{2 cm}
%
%%Begin InstantTeX Picture
\let\picnaturalsize=N
\def\picsize{3in}
\def\picfilename{loop.eps}
%If you do not have the picture file add:
%\let\nopictures=Y   
%to the beginning of the file.
\ifx\nopictures Y\else{\ifx\epsfloaded Y\else\input epsf \fi
\let\epsfloaded=Y    
\centerline{\ifx\picnaturalsize N\epsfxsize \picsize\fi
\epsfbox{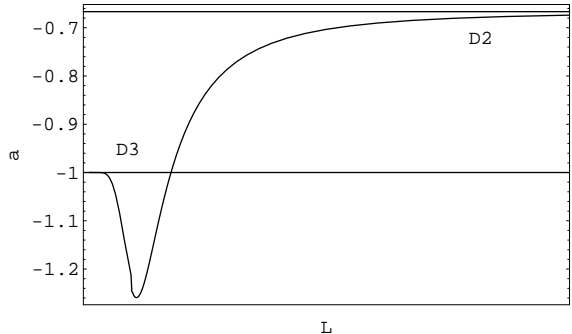}}}\fi
%%End InstantTex Picture   
\caption{{\small $a({\cal L})$ in terms of ${\cal L}$. Here we see how the finite size 
effects interpolate between the D$2$ and D$3$ branes.}}  
\label{loop1}
\end{figure} 

In the figure (\ref{loop1}) we have plotted our result. We have normalized
the variables in the integrals, working with dimensionless parameters. The
election has been such that the results can be written in terms of
${\cal E}=E_{qq}\tilde{R} $ as a function of ${\cal L}=L\tilde{R}^{-1}$.

Assuming the usual functional form 
\be
{\cal E}\sim {\cal L}^{a({\cal L})}
\ee
we can use the variable $a({\cal L})$ to see how the quark-antiquark potential goes 
from the unwrapped D$3$-brane behavior, $a({\cal L})=-1$, to the expected for the 
D$2$-brane in a decompactified background, where  $a({\cal L})=-2/3$ \cite{son2}. The complete 
result is shown in (\ref{loop1}). We see that when the distance between quarks is large
 enough compared with the compactification radius, $ L\gg R $, the system feels a small 
compact dimension so the propagator of gauge fields going from one quark to the other 
does not include any KK mode. 
In this  case we approach the T-dual system, that is the D$2$-brane. When the radius 
is large, or the quarks are very close to each other, $ R\gg L$, the KK modes are very 
light and all of them must be taken into account. Then the system approaches the 
unwrapped  D$3$-brane one.

Let us now explore the other loop geometry we are interested in. Apart from the 
spatial separation between quarks, one can consider that
they have a different `flavour', that is, that, taken as vectors in the
R-symmetry space $S^1\times S^5$, the quarks point at different directions.
This affects the Wilson loop and, therefore, the potential. We will
concentrate on phase differences in the circle, which is the new parameter
here. The angular difference on the sphere \cite{maldaloop,rey} does not play any 
role in the T-duality map.
The way to calculate in this case is the same as before, but adding
some new terms. Again, we have to compute the area of the world-sheet, which
is given by the Nambu-Goto action. It is such that
\ber
X^i(x,\tau)=x L \nonumber \\
\theta(1,\tau)-\theta(0,\tau)=\theta \\
T=\tau
\eer
and the configuration is static so nothing depends on $\tau$. The form of
the world-sheet in the target space is the same as before, except that the
spatial sides of the rectangle are not exactly in the $i^{\mbox{\scriptsize th}}$
direction but in a general direction in the $(X^i,\theta)$ plane. The action
is given by
\be
S=\frac{1}{2 \pi \alpha'}\int dx \sqrt{G_{x x} G_{\tau \tau} +G_{u u} G_{\tau \tau} 
(\partial_x u)^2 +G_{\tau \tau} G_{\theta \theta} (\partial_x \theta)^2}
\ee

   We would like to compare this expression to its T-dual. In order to do
that, we have, firstly, to look for a configuration that is the T-dual of
this one. It is important to notice that the dual magnitude to an angle in
one circle is not another angle in the dual circle because that would mean
that movements, and therefore momenta in both circles would be related by
the duality, which is not true. A string stretched between two points placed
a distance apart in the circle of the D2-brane solution has an energy
related to its fractional winding number; so we expect that the T-dual
configuration has some fractional momentum in the IIB supergravity. From the gauge 
theory point of view, the R-symmetry separation of quarks on the D$2$-brane should 
be represented by a pair of quarks, living on the T-dual D$3$-brane, moving with 
different momenta in the compact direction.

The main difference of the string worldsheet that describes the Wilson loop in 
supergravity, is that the compact scalar must have momentum so instead of the 
static configuration we should have
\ber
X^i(x,\tau)=x L \nonumber \\
\psi(x,T_{\mbox{\scriptsize max}})-\psi(x,0)=\psi \\
T=\tau
\eer
Now the action is
\be
S=\frac{1}{2 \pi \alpha'}\int dx d\tau \sqrt{G_{x x} G_{\tau \tau} 
+G_{u u} G_{\tau \tau}
(\partial_x u)^2 +G_{\tau \tau} G_{\psi \psi} (\partial_\tau
\psi)^2+G_{u u} G_{x x}(\partial_x u)^2 (\partial_\tau \psi)^2}
\ee

Both actions are quite different. One integrates fields in one dimension and
the other in two, besides, the second has a term which is quartic in the
velocities. 
Here we will show that both actions describe the dynamics of T-dual systems. The
procedure we will use assumes that, similar to what happens in the string sigma model 
\cite{lozano}, these two actions are related by a canonical transformation. 
 
The Hamiltonians of the systems are simply computed to give
\be
H_{D2}=\sqrt{G_{x x} G_{\tau \tau}+\frac{G_{x x}}{G_{\theta \theta}} P_\theta^2 
+\frac{G_{x x}}{G_{u u}} P_u^2}
\ee
and
\be
H_{D3}=\sqrt{G_{x x} G_{\tau \tau}- G_{x x} G_{\psi \psi} (\partial_\psi
\psi)^2+ \frac{G_{x x}}{G_{u u}} P_u^2}
\ee
Now we simply see that the canonical transformation should be
\be
P_\theta=-i\partial_\tau \psi 
\label{canonical}
\ee
with no change in the coordinates and momenta. There is a small difference between the 
transformation in (\ref{canonical}) and those in \cite{lozano}. Here we simply adopted 
the rule in  \cite{lozano} to the Euclidean case. Imposing the equality of the
 Hamiltonians $H_{D2}$ and $H_{D3}$ we recover the Buscher's transformation rule for 
the metric
\be
G_{\psi\psi}=\frac{1}{G_{\theta \theta}}
\ee
 that finally shows that the worldsheet configurations we described below are T-dual.

\section{Conclusions}

In this work we have studied the dynamics of D-branes sitting on backgrounds
with toroidally compactified dimensions. Concretely we focused our analysis
on the D$3$ and D$2$-branes with one compactified dimension. All our results
are straightforwardly extensible to more general situations. We worked in the
framework of the Maldacena duality, trying to clarify how T-duality enters 
into the holographic conjecture.

Our principal interest has been the analysis of finite size effects on the
dynamics of the systems. We studied their influence on the dynamics of the
brane and then, by T-duality, we showed how the possible corrections appear
in the dual system.

The analysis started by a detailed study of the background geometry of a
D$2$-brane with a compactified transverse direction. In order to describe
the effect of the compact circle we used the multicentered solution of
IIA supergravity. We clearly explained how the near horizon limit has to be
taken. The background is described by a series of  harmonic functions, the
eq.(\ref{series}), expressing the presence of an infinite tower of winding
modes. We found a transition point, ${\cal R}\sim u_{\bot}$,  where the
accuracy of any truncation of the series is ever worse. Physically this
means that at this point finite size effects become very important. It is 
then preferable to use a Poisson resummation formula for the series. It does 
not mean that we are making a T-duality transformation. In fact the system 
now behaves as a continuous distribution of branes along the compact direction, 
still staying in type IIA string theory.

We are, in some sense, forced to make use T-duality and go to the three-brane
system when the physical radius, $R_{\mbox{phys}}$, is smaller than the string
scale. We showed that this point is always reached at values of $u_{\bot}$
larger than ${\cal R}$, so the system has passed through the continuous
distribution-phase described above. On the other hand we see that at these
energies in the T-dual system finite size effects are irrelevant.

In sections 3 and 4 we present what can be learned about the gauge theory 
from supergravity. We know that the coordinate dependence of the expectation
value of background fields corresponds to the renormalization group flow 
of the corresponding quantities in the gauge theory. We saw that in our cases
the renormalization results coming from supergravity appear in mass-dependent 
renormalization scheme. It is due to the fact that, in order to obtain information 
from finite size effects, we should include all the infrared degrees of freedom 
of the theory. Finally we showed that the angular separation of a test 
D$2$-brane from the $N$-branes source is mapped on the D$3$-brane system 
as a Wilson Line. In this case the test object sees a constant gauge field on the 
source.    
 
The AdS/SYM allows the description of the strong 't Hooft coupling of the
gauge theories when the string coupling is small. In the case of non-conformal
brane configurations this establishes some limits on the phase space describable in
terms of supergravity. In our work we dealt with two non-conformal systems.
In the case of the D$2$-brane the renormalization is shown by the running of the string
coupling and of the size of the physical radius of the compact dimension. The
latter reflects the change of the masses of winding modes. The
D$3$-brane case is more subtle. In order to see the running of the
coupling one cannot use the ten-dimensional string coupling but one must use
the coupling of the T-dual D$2$-brane. The explanation of this requirement
could be seen from pure gauge theory arguments. When we compactify the one
of the three-brane coordinates and express our fields in terms of KK modes in
two dimensions we really deal with the D$2$-brane which coupling constant is the
T-dual of the initial one. Another way to see the non-conformal nature of
the wrapped three-brane, using fields in three dimensions, is looking at the
energy dependence of their masses.

Finally we studied the dynamics of Wilson loops in the systems. We presented two
possible configurations of the loop. They correspond to two different geometries 
of the string worldsheet that describe the quark-antiquark interaction from 
supergravity.

The simplest one described a pair of quarks at a distance $L$ in the compact direction
and with the same position in the angular directions. This configuration 
corresponds to a static string worldsheet. We studied the evolution of
the  properties of the system in terms of the supergravity parameters. We saw how 
the system goes from a pure D$2$-brane loop to the three-brane one. Our computation
allowed us to study how the presence of a compact direction could affect the system.

The other quark-antiquark configuration we considered, includes an angular distance,
on the circle, between infinite strings describing static quarks of the two-brane.
 In this case we showed that T-duality converts this static system into a 
time-dependent one on the three-brane. Concretely it corresponds to quarks moving on 
the compact worldvolume direction with different momenta. We constructed this configuration
from physical arguments and finally showed that it is exactly the same as the initial 
one but expressed in canonical transformed variables.  

\section{Acknowledgments}

We are very grateful to  M. Su\'arez, M.A.R. Osorio, J. F. Barb\'on, P. Silva 
 A. Nieto and V. Di Clemente for 
enlightening conversations. The work of M.L.M. is supported by a M.E.C. grant under 
the FP97 project.

\end{document}